# Stochastic resonance in a metal-oxide memristive device

A.N. Mikhaylov[a,*], D.V. Guseinov[a], A.I. Belov[a], D.S. Korolev[a], V.A. Shishmakova[a],
M.N. Koryazhkina[a], D.O. Filatov[a], O.N. Gorshkov[a], D. Maldonado[b], F.J. Alonso[c], J.B. Roldán[b],
A.V. Krichigin[a], N.V. Agudov[a], A.A. Dubkov[a], A. Carollo[a,d], B. Spagnolo[a,d,e]

[a] *Lobachevsky University, 23/3 Gagarin prospect, 603022 Nizhny Novgorod, Russia*
[b] *Departamento de Electrónica y Tecnología de los Computadores, Universidad de Granada, Avd. Fuentenueva S/N, 18071, Granada, Spain*
[c] *Departamento de Estadística e Investigación Operativa, Universidad de Granada, Avd. Fuentenueva S/N, 18071 Granada, Spain*
[d] *Dipartimento di Fisica e Chimica Emilio Segrè, Group of Interdisciplinary Theoretical Physics, Università di Palermo, Viale delle Scienze, Edificio 18, I-90128 Palermo, Italy*
[e] *Istituto Nazionale di Fisica Nucleare, Sezione di Catania, Via S. Sofia 64, 95123 Catania, Italy*

## Abstract

The stochastic resonance phenomenon has been studied experimentally and theoretically for a state-of- art metal-oxide memristive device based on yttria-stabilized zirconium dioxide and tantalum pentoxide, which exhibits bipolar filamentary resistive switching of anionic type. The effect of white Gaussian noise superimposed on the sub-threshold sinusoidal driving signal is analyzed through the time series statistics of the resistive switching parameters, the spectral response to a periodic perturbation and the signal-to- noise ratio at the output of the nonlinear system. The stabilized resistive switching and the increased memristance response are revealed in the observed regularities at an optimal noise intensity correspond- ing to the stochastic resonance phenomenon and interpreted using a stochastic memristor model taking into account an external noise source added to the control voltage. The obtained results clearly show that noise and fluctuations can play a constructive role in nonlinear memristive systems far from equilibrium.

*Keywords:*
Memristor  Resistive switching, Yttria-stabilized zirconium dioxide  Tantalum oxide, Time series statistical analysis, stochastic  model, Stochastic resonance.

## 1. Introduction

Thin-film devices with the effect of resistive switching (RS) [1] attract strong attention from researchers and engineers in connection with the prospects of their use as a new element base for non-volatile memory devices, that is resistive random access memory (RRAM), within in-memory and neuromorphic computing systems, as well as neurohybrid systems. They are designed not only to solve fundamental problems of traditional architectures of information and computing systems, but also to take the next step towards their symbiosis with biological (neuronal) systems [2-13]. From this point of view, the most attractive and CMOS-compatible metal-oxide devices exhibit the effect of bipolar RS, which consists in a reversible change in the device resistance under the action of an electric voltage of opposite sign exceeding a certain threshold. The generally accepted mechanism of such RS is associated with the destruction and recovery of conducting filaments (CFs) formed due to the migration of oxygen vacancies/ions in the oxide layer during electroforming [14-15]. A breakthrough in this direction occurred in 2008 [16], when the RS effect in metal oxides was correlated with the definition of a memristor, as a memory resistor proposed by Leon Chua back in 1971 [17]. A generalized definition of the memristor as a nonlinear dynamical system [18] brings the analysis of the phenomena responsible for RS at the macroscopic level of the system, which makes it possible to use the rich arsenal of nonlinear physics methods for the interpretation and control of the RS.

Despite significant progress in the physics and technology of memristive devices, their widespread practical application is constrained by insufficient stability, high variability of RS parameters, lack of understanding of the drift-diffusion processes as well as their degradation during operation [19-23]. A combination of diverse transport phenomena under the influence of concentration gradients, electric field and temperature, as well as redox reactions, is responsible for the CF reconstruction even in the simplest metal-oxide devices. These physico-chemical phenomena occur on different time scales, out of equilibrium conditions, and with intrinsic random fluctuations, which give rise to the inherently stochastic nature of the RS process [24]. Traditional approaches to increasing the stability of RS parameters include memristive device engineering by selecting the optimal stacks of oxide and electrode materials [25] and the use of multilayer structures [26,27], as well as special programming techniques [28,29]. These approaches are quite effective for demonstration purposes in the implementation of small-scale prototypes of computer systems [2,15], however, technology and circuit-level overheads question the prospects for further scaling.

Relatively new approaches to control the dynamic response of memristors based on the use of attractors [30], bifurcations [31] and deterministic chaos [32,33] have not yet been implemented experimentally. The main reason for this is that an RS-based memristive device is not a deterministic dynamical system, but a complex stochastic system, the description of which requires taking into account the noise and nonlinear relaxation phenomena inherent in real systems. Therefore, to analyze these new approaches, it is necessary to use stochastic memristor models.

In Ref. [34], such a stochastic model is proposed in which the random jumps of microscopic defects along the lattice sites are described by a smooth coarse-grained function of the defect concentration. The concentration of the coarse-grained defect determines the average resistance value of the memristor at any time. The Fokker-Planck equation (FPE) for the defect concentration involves the height of the activation energy for the defect jumps, the intensity of thermal fluctuations, and a number of other parameters that characterize the material properties of memristors, including interfaces with electrodes. In this work, we demonstrate that this model also allows for an external noise source added to the control voltage to be taken into account.

Recent studies have convincingly shown that, in nonlinear systems, the effect of noise can induce new, more ordered regimes that lead to the formation of more regular structures, increase the degree of coherence, cause an increase in gain and signal-to-noise ratio (SNR) at the output of the system, etc. In other words, fluctuations can play a constructive role in nonlinear systems far from equilibrium [35-42].

Among the phenomena characterized by the constructive role of noise and constituting the subject of a new rapidly developing field of investigation [43,44], it is possible to distinguish the stochastic resonance (SR) observed in different nonlinear systems under the simultaneous presence of noise and a periodic signal [45]. Under some conditions, an increase in the intensity of internal or external noise leads to a coherent response of the nonlinear system, with a
non-monotonous behavior and a maximum of the SNR as a function of the noise intensity.

To detect the SR, it is necessary to have a threshold-type non-linearity, periodic perturbation, and a source of white or colored noise in the system under study. These conditions can be realized for metal-oxide memristive devices. The filament-type RS process is nonlinear and has pronounced threshold properties. The standard endurance test is carried out by applying periodic bipolar pulse trains that meet the requirements of a coherent driving signal. The imposition of noise on such a signal at its sub-threshold amplitude can lead to the stabilization of RS and many other useful effects associated with the constructive role of noise (e.g. stochastic resonant activation [46-48]).

This innovative approach to solve the problem of increasing the stability of memristors, based on the use of the SR phenomenon, was proposed for the first time in the theoretical work of Ref. [49]. A dynamical memristor model was taken as a basis and noise was added as simply as possible in order to qualitatively demonstrate the occurrence of the SR phenomenon. In the experimental work of Ref. [50], a white Gaussian noise was superimposed on the rectangular switching signal. An increase in the temporal stability of the memristor was noted in the case of switching with superimposed noise. The effect was most pronounced at certain noise intensity, indicating the possibility of the SR manifestation in memristive devices. At the same time, it was shown that the simplest model proposed in Ref. [49] does not fit the experiment if an external noise is used as a source of fluctuations, that is a noise signal added to the externally applied voltage. The model proposed in [50] was also based on a dynamical model, adding external noise to the driving signal. Consequently, the authors obtain a stochastic model containing a nonlinear multiplicative source of noise. The numerical simulations of this stochastic model show a good agreement with the experiments, while the theoretical analysis is possible only under the assumption of small noise intensity. Preliminary experimental data on the influence of external noise on the RS driven by a sinusoidal signal have been recently reported [51], and the authors continue to explore some positive aspects of this effect, for example, the mitigation of the bit error rate problem in RRAM crossbars [52]. However, to date and to the best of our knowledge, there is no information in the literature on more in-depth studies of the SR phenomenon in memristive devices both in terms of RS parameters improvement and basic SR regularities.

In the present work, such a study has been carried out for integrated memristive microdevices, which are made on the basis of an optimized multilayer structure consisting of zirconium oxide and tantalum oxide [27] and which demonstrate reproducible switching with good retention and endurance. These devices have worked well in the study of the dynamics of the resistive state under the influence of noise [53] and served as a basis for the development and calibration of a large-scale stochastic model [34]. Since a reliable demonstration of the influence of noise superimposed on a periodic switching signal on RS parameters requires the analysis of long time series, the method of time series statistical analysis (TSSA) has been used for this purpose, which was proved valid in the study of various memristive devices [54].

The observed regularities in the behavior of RS parameters, such as the ratio of resistances in the high resistance state (HRS) and the low resistance state (LRS), a variation of resistive states and switching voltage, have been compared with the characteristics traditionally studied in the analysis of SR phenomena, such as the spectrum of the response of the nonlinear system to a periodic perturbation and the SNR at the output of the system as a function of the intensity of the input noise.

Device fabrication and measurement setup are described in Section 2, the stochastic model will be introduced in Section 3 and the main results will be presented and discussed in Section 4, including the time series analysis of the measurements. Finally, we will summarize the conclusions in Section 5.

## 2. Device Fabrication and Stochastic Resonance Experiment

The memristive device under study was fabricated on the top of industrial TiN (25 nm) / Ti (20 nm) metallization deposited on the oxidized $SiO_2$ (500 nm) / Si substrate by magnetron sputtering of high-purity Ti target. The $Ta_2O_5$ (10 nm) and $ZrO_2$(Y) (10 nm) films were deposited by the method of radio-frequency magnetron sputtering of Ta and $ZrO_2(12\%Y_2O_3)$ targets, respectively, in the mixed Ar (50%) and $O_2$ (50%) atmosphere at the substrate temperature of 300 °C. The Ta (8 nm) and Au (20 nm) top layers were deposited by the method of direct-current magnetron sputtering of high-purity metal targets in Ar atmosphere at the substrate temperature of 200 °C. According to the previous paper [27], this combination of materials and deposition parameters provides robust RS due to the presence of grain boundaries in $ZrO_2$(Y) as preferred sites for CF nucleation, self-assembled nanoclusters as field concentrators in the $Ta_2O_5$ film, and oxygen exchange between the $Ta_2O_5$, $ZrO_2$(Y) layers and interface with the bottom TiN electrode. The present experiments were performed on isolated cross-point devices with a size of 20 μm × 20 μm fabricated on a memristive test microchip mounted in a standard 64-pin package.

A classical experiment to confirm the SR phenomenon [55], [56], [57], [58] consists of the application of a periodic driving signal to a nonlinear system with an amplitude that is insufficient to cause a regular periodical response of the system to the periodic signal. As the noise level added to the driving signal increases, the system response becomes regular and, when the optimum noise level is reached, the maximum SNR is realized. A further increase in noise intensity produces a loss of coherence of the system output and the dynamical behavior is strongly controlled by the noise.

To conduct such experiment, an external Gaussian noise signal was generated using the ADSViewer-2 (v.015) pseudorandom number generator [59] and characterized by the voltage amplitude (maximum variation in voltage values). This noise signal was superimposed on a sinusoidal voltage signal, the period of which was 0.1 s and the peak-to-peak amplitude was varied in the range of 0.5–2.0 V to find the sub-threshold switching mode. Typical input voltage signals are shown in Fig. 1 and consist of a sine waveform of 1 V amplitude combined with a Gaussian distributed noise source with amplitudes of 0.1 V and 1 V. The current response of a memristor was taken from a load resistor (100 Ω) connected in series with the memristor. DAC and ADC were used to generate the input signal and register the output as part of the USB 6211 multifunctional device (National Instruments ™). The sampling rate was 6.553 kHz corresponding to the correlation time of $1.526 \cdot 10^{-4}$ s. The output signal power spectrum was calculated by the fast Fourier transform technique for the time series of selected parameter.

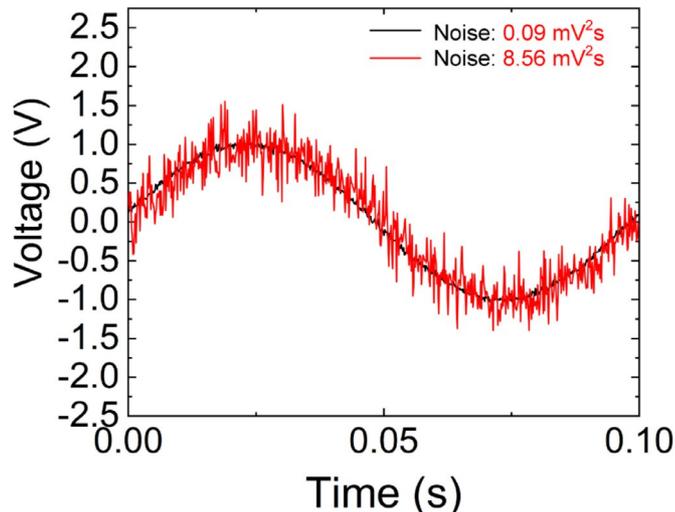

**Fig. 1.** Input voltage vs. measurement time for a single cycle applied to the mem- ristive device under study for two noise intensities.

response of a memristor was taken from a load resistor (100 ) connected in series with the memristor. DAC and ADC were used to generate the input signal and register the output as part of the USB 6211 multifunctional device (National Instruments TM). The sampling rate was 6.553 kHz corresponding to the correlation time of 1.526•10−4 s. The output signal power spectrum was calculated by the fast Fourier transform technique for the time series of selected parameter.

**3. Stochastic Memristor Model**

To describe the observed stochastic RS behavior, a recently proposed stochastic memristor model [34] was used. The model is based on the description of a random hopping of dielectric structural defects (oxygen vacancies), which are positively charged, between the trapping sites within the structure of the dielectric material. These elementary random events provide the diffusion process that leads to the formation or destruction of CF. The diffusing particle motion is described by the following Langevin equation

$$\frac{dx}{dy} = -\frac{\partial U(x,V)}{\partial x} + \xi(t), \quad (3.1)$$

where $x$ is the coordinate of a particle diffusing between electrodes located at $x = 0$ and $x = L$, $U(x,V)$ is the potential profile defining the regular force acting on the particle and $V$ is the potential difference between electrodes. The potential profile $U(x,V)$ is represented by the periodical part $\Phi(x)$ describing the potential wells separated by the barriers in periodical structure of oxide dielectric and the force of the external electric field $F$

$$U(x,V) = \Phi(x) - Fx, \quad (3.2)$$

where $F=qV/\varepsilon L$, L is the distance between electrodes, q is the charge of particle and ε is the dielectric constant. The height of barriers in the periodic term $\Phi(x)$ is the activation energy $E_a$, needed for a particle to surmount the barrier and move to the neighboring well in a random direction (see [34] for more details).

The random force $\xi(t)$ is a white Gaussian noise with $\langle \xi(t) \rangle = 0$ and $\langle \xi(t)\xi(t+\tau) \rangle = 2\theta_\xi \delta(\tau)$, where $\delta(\tau)$ is the delta function and $2\theta_\xi$ is the noise intensity. When the fluctuations have only thermal nature, the intensity is proportional to the temperature $\theta_\xi = k_B T$, where T is the temperature and $k_B$ is the Boltzmann constant.

In the present paper, we apply also an external noise source to the system, and the driving voltage reads

$$V(t) = V_0 + \zeta(t), \quad (3.3)$$

where $V_0$ is the deterministic part of the potential difference and $\zeta(t)$ is another white Gaussian noise source with $\langle \zeta(t) \rangle = 0$ and $\langle \zeta(t)\zeta(t+\tau) \rangle = 2\theta_\zeta \delta(\tau)$. The intensity of the external noise is $2\theta_\zeta$.

By inserting (3.3) into Eqs. (3.1) and (3.2) we will obtain the Langevin equation with two additive white Gaussian noise sources which can be replaced by an equivalent one

$$\frac{dx}{dy} = -\frac{\partial U(x,V_0)}{\partial x} + \nu(t), \quad (3.4)$$

where $\langle \nu(t) \rangle = 0$ and $\langle \nu(t)\nu(t+\tau) \rangle = 2\theta_\nu \delta(\tau)$ and the value $\theta_\nu$ reads

$$\theta_\nu = \frac{q^2}{\varepsilon^2 L^2} \theta_\zeta + \theta_\xi. \quad (3.5)$$

The Eq. (3.4) is the same as (3.1) but with a different value for the noise intensity due to the addition of external noise. Thus, in the framework of this model we may consider the external noise as an addition to the thermal noise, and in this sense its influence is equivalent to the effect of temperature (i.e. heating or cooling of the device).

According to Ref. [34] and basing on Eq. (3.1), or now (3.4), we can obtain the FPE for the smoothed coarse-grained concentration of defects n(x,t)

$$\frac{\partial}{\partial t} n(x,t) = \frac{\partial}{\partial t}\left[ n(x,t) \frac{\partial U_{eff}(x,V)}{\partial x} \right] + D_{eff} \frac{\partial^2}{\partial x^2} n(x,t), \quad (3.6)$$

with the following effective potential profile,

$$U_{eff}(x,V) = -v_{eff}\, x, \quad (3.7)$$

and effective drift and diffusion coefficients

$$v_{eff} = \frac{2l}{\tau_{kr}} \sinh \alpha, \quad (3.8)$$

$$D_{eff} = \frac{l^2}{\tau_{kr}} \cosh \alpha, \quad (3.9)$$

where $l$ is the period of the periodical part $\Phi(x)$ of the potential profile, $\alpha = BV_0/2\theta_\nu$, where $B = ql/\varepsilon L$ is the fitting coefficient and $\tau_{kr}$ is the Kramers time

$$\tau_{kr} = \tau_0 \exp \beta, \quad (3.10)$$

where $\beta = E_a/\theta_v$ is the dimensionless value of potential barriers for diffusing particles when $V_0 = 0$, and $\tau_0$ is the prefactor. When a deterministic, or mean, value of voltage $V_0 \neq 0$ is applied, the potential barriers decrease in one direction and increase in opposite direction by the value $\alpha$. As an example, a dimensionless potential profile for $V_0 > 0$ is shown in Fig. 2, where the potential barriers decrease in the positive direction of $x$ down to $\beta - \alpha$ and appropriately increase in the negative direction up to $\beta + \alpha$. The nonlinear dependence of the effective drift coefficient (3.8) on driving voltage $V_0$ provides the threshold property to the model.

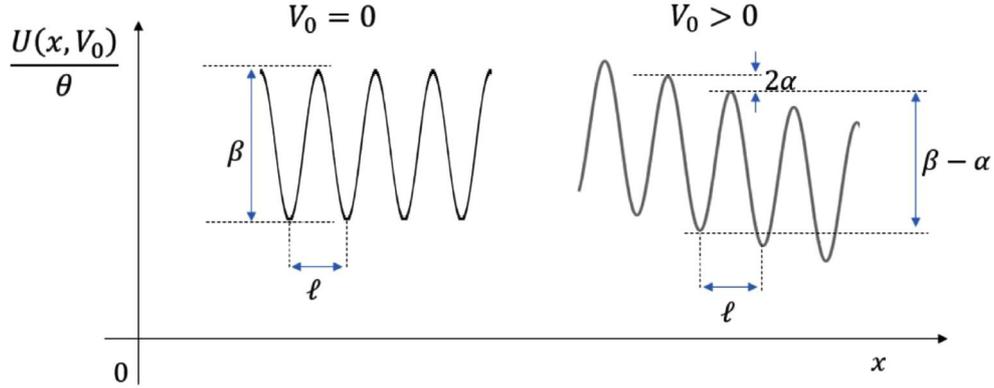

**Fig. 2.** Dimensionless potential profile (3.2) for $V_0 = 0$ and $V_0 > 0$, where $\beta = E_a/\theta_v$ is the dimensionless height of potential barrier under $V_0 = 0$ and $\alpha = BV_0/2\theta_v$ is the value of barrier height variation under $V_0 \neq 0$.

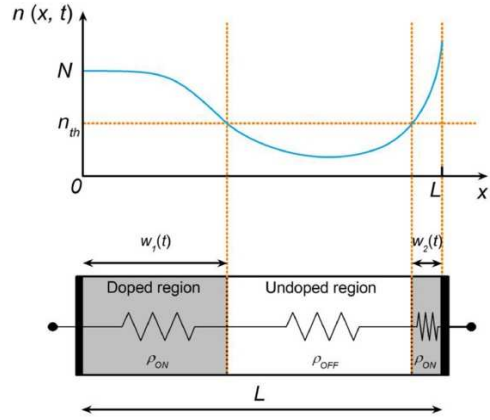

**Fig. 3.** Diagram showing the smoothed coarse-grained concentration of defects $n(x, t)$ with the equivalent electric circuit of the memristor model.

The boundary conditions for FPE (3.6) should be specified according to the material of electrodes. In our case we consider relative concentration of defects N=100% at the easily oxidizable TiN/Ti electrode

$$n(0, t) = N ,\qquad(3.11)$$

and reflecting boundary conditions for the inert Au electrode (with a thin adhesive Ta sublayer), which assume that the flow of diffusing particles through this electrode equals zero

$$G(L, t) = v_{eff} n(L, t) - D_{eff} \left.\frac{\partial n(x,t)}{\partial x}\right|_{x=L} = 0. \qquad(3.12)$$

The value of resistivity $\rho(x, t)$ in each point of $x$ and time $t$ is defined by a nonlinear dependence $\rho(n)$. In the present manuscript we use a threshold-like function $\rho(n)$ (see also Fig. 3):

$$\rho(x,t) = \rho(n) = \begin{cases} \rho_{ON}, & n > n_{th} \\ \rho_{OFF}, & n < n_{th} \end{cases}. \qquad(3.13)$$

The threshold value is assumed as follows: $n_{th}$ = N/2. Note, that in the presented model the function $n(x)$ can be non-monotonous as shown in Fig. 3, and this is related to the effect of transient bimodality [60]. This means that the CF is formed from two sides. The total resistance of memristor (memristance) $R_m(t)$ is equal to the following integral

$$R_m(t) = \int_0^L \rho(x,t)dx . \qquad(3.14)$$

The other parameter values are the following: $\frac{l^2}{\tau_0} = 6 \cdot 10^{-13}$ cm²s⁻¹, L = 20 nm, $\beta = 25$ when $\theta_v = \theta_\xi$, that is in the absence of external noise.

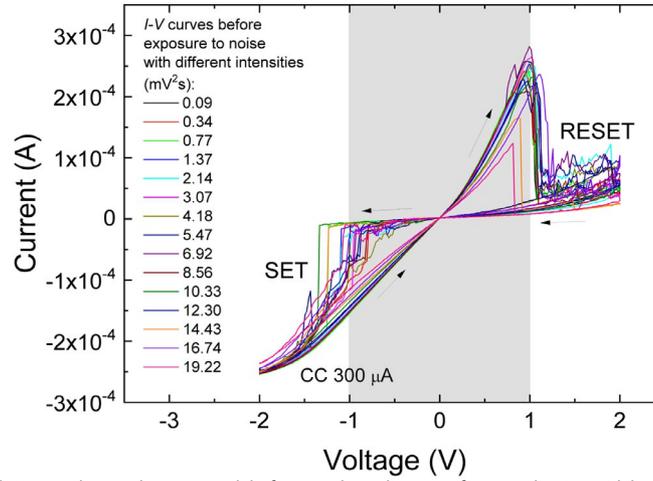

**Fig. 4.** *I-V* characteristics of the memristive device under study measured before each application of a periodic sinusoidal signal with the amplitude of 1 V and dif- ferent noise intensities.

## 4. Results and Discussion

### 4.1. Resistive Switching Regularities

In Fig. 4 are shown the typical *I-V* characteristics of the memristive device under study in the bipolar RS mode measured before the application of a periodic signal with different noise intensities. After each application, these measurements were used to check and confirm the reproducibility of memristive device characteristics. As can be seen in Fig. 4, the transition to HRS (RESET process) occurs with positive bias, and the relatively abrupt reverse transition to LRS (SET process) occurs with negative bias. The figure also shows that due to the internal asymmetry of these processes, the SET process is performed at a lower voltage than the RESET process and requires the current compliance to avoid damage of the device.

To select a sub-threshold signal level, the response of the memristive device to different amplitudes of the driving sinusoidal voltage signal was studied first. Before each measurement, the memristor was switched to its initial HRS. At amplitudes lower than 0.8 V, RS was not revealed, and, at amplitudes greater than 1.5 V, reproducible switching between the extreme resistive states was observed.

To select a sub-threshold signal level, the response of the memristive device to different amplitudes of the driving sinusoidal voltage signal was studied first. Before each measurement, the memristor was switched to its initial HRS. At amplitudes lower than 0.8 V, RS was not revealed, and, at amplitudes greater than 1.5 V, reproducible switching between the extreme resistive states was observed.

The evolution of *I-V* characteristics with external noise intensity under the constant amplitude 1 V of the sinusoidal driving signal is shown in Fig. 5. The left panels of Fig. 5 represent the experimental data, and the right panels are simulated according to the coarse-grained stochastic model. The model provides a value of memristance $R_m(t)$ averaged over an infinite ensemble of systems. The appropriate *I-V* characteristic does not contain fluctuations and corresponds to the average value of $R_m(t)$. The experimental results are measured on a single system and are averaged over a restricted number of switching circles. Therefore, the experimental *I-V* characteristics are noisy. In the absence of external noise, the amplitude of the sinusoidal signal was insufficient for the complete and regular SET and RESET processes, as is shown in Figs. 5a and 5b.

If we add a small amount of noise ($\theta_\zeta$ = 0.09 mV$^2$s), the RS starts to occur, and small hysteresis appears on *I-V* characteristics (see Fig. 5c and 5d). If we continue to raise the noise intensity, we observe a further increase in the amplitude of the corresponding periodical variation of $R_m(t)$. As a result, the area of the *I-V* hysteresis loop grows (see Fig. 5e and 5f).

With further noise intensity increase, RS occurs at values of the driving voltage lower than the maximal value 1 V. As one can see in Fig. 5g – 5j, the greater the noise intensity the smaller the switching voltage. In such a way, the area of the hysteresis loop starts to decrease and disappears for high enough noise intensity.

This variation of RS properties can be explained, when we take into account the changing difference between the two time scales: the average switching time of memristive device $\tau$ and the period of external driving $T$. When $\tau > T$, the RS does not occur. This is a sub-threshold mode. The switching appears, when $\tau \leq T$. In this case we start to observe the hysteresis. When the switching time is very short compared to the period of driving signal ($\tau \ll T$), the device switches to a new state very quickly compared to the rate of voltage variation. That means for $\tau \ll T$ the hysteresis is small or absent while the switching occurs. Therefore, the hysteresis exists only for intermediate case, when $\tau \leq T$, and disappears when $\tau > T$ or $\tau \ll T$.

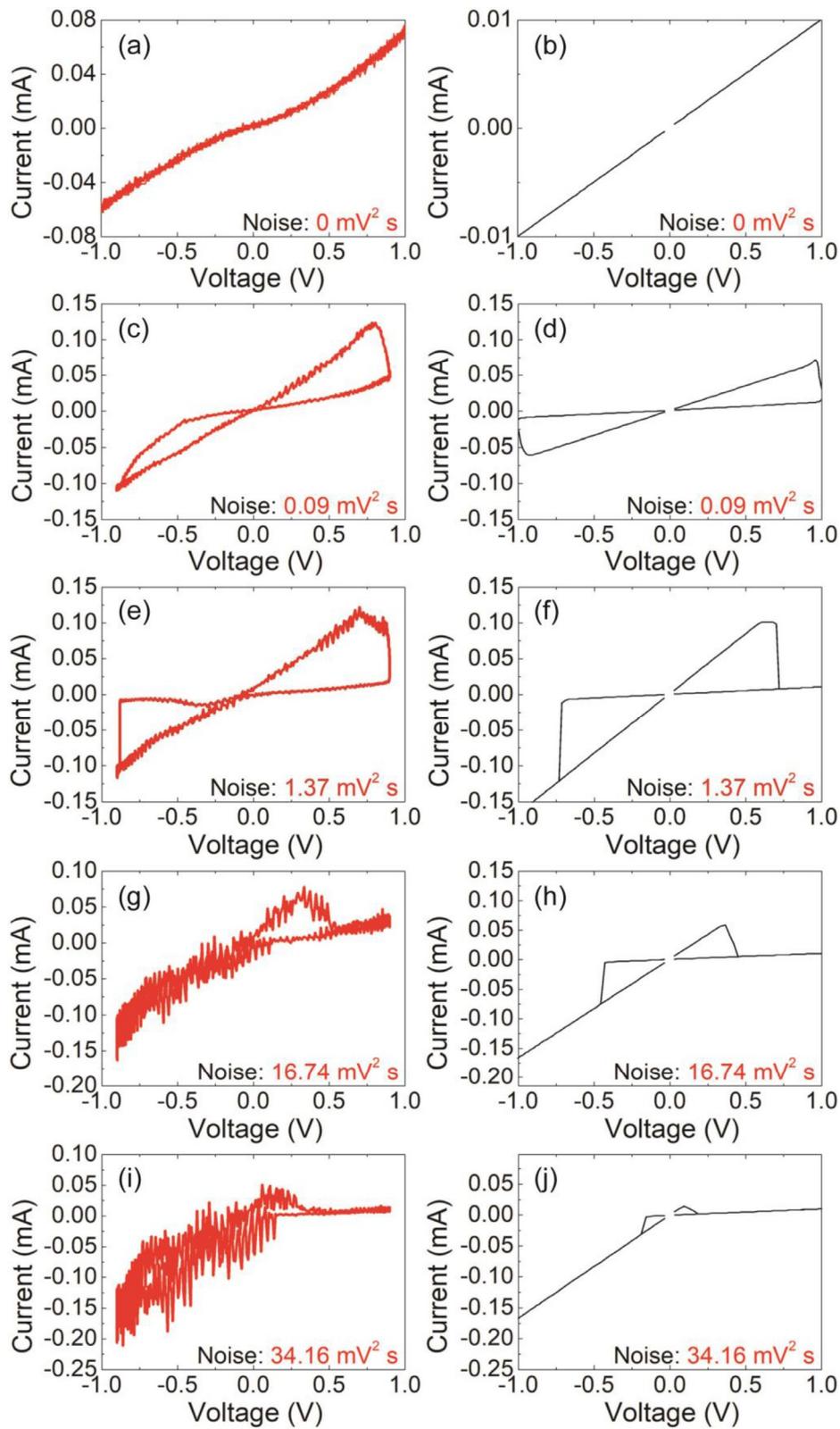

**Fig. 5.** I-V characteristics of the memristive device under study measured (left panels) and simulated (right panels) under the amplitude of driving signal 1 V and frequency 10 Hz for the growing intensity of external noise.

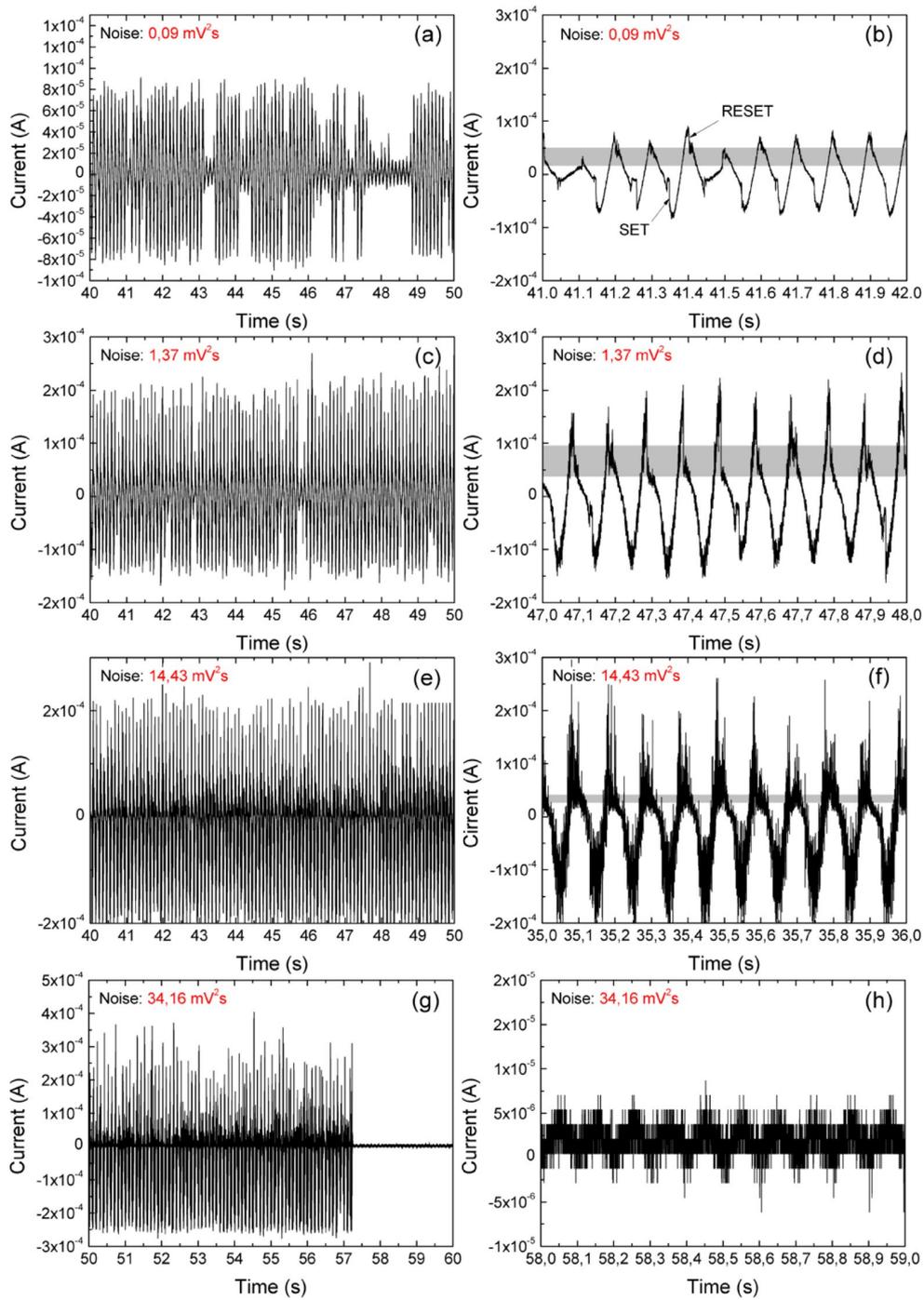

**Fig. 6.** Time series of the current response of memristive device under study on different time scales, measured for different noise intensities added to a sinusoidal driving signal with amplitude of 1 V.

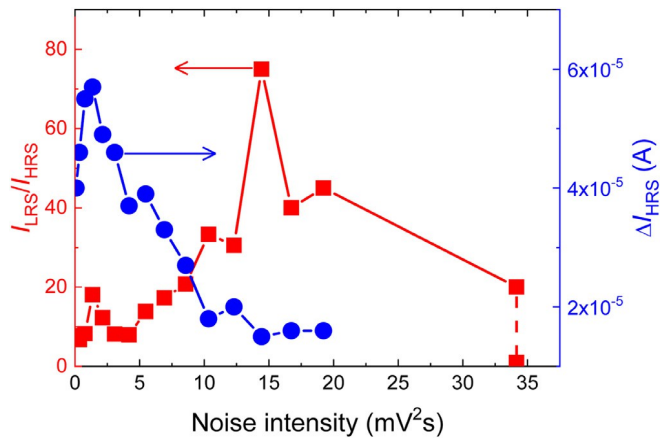

**Fig. 7.** The LRS and HRS current ratio ($I_{LRS}/I_{HRS}$), as well as the variation of current after switching to HRS ($\Delta I_{HRS}$) vs. the noise intensity.

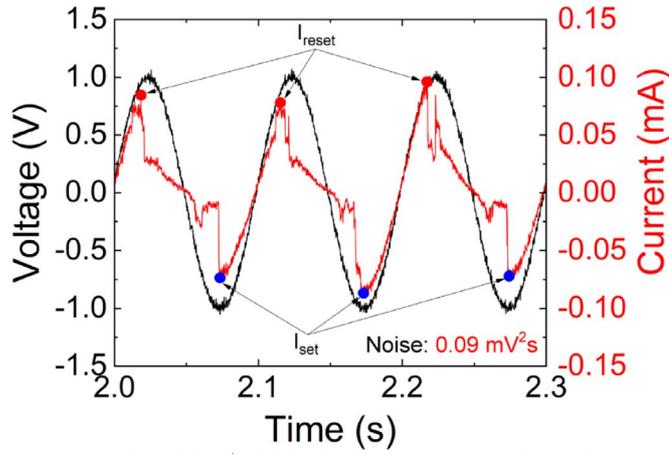

**Fig. 8.** Experimental current and voltage versus time elapsed (0.1s/cycle) for three consecutive RS cycles in a long series for the memristive device under study for a noise intensity of 0.09 mV²s.

According to the coarse-grained stochastic model [34], the switching time (or relaxation time to the stationary state) decreases with voltage as well as with intensity of fluctuations. Therefore, if we select a small enough, sub-threshold amplitude of driving voltage providing $\tau > T$, the device will not switch in average as shown in Fig. 5a and 5b. In order to provide RS under the fixed parameters of driving, we can reduce the switching time $\tau$ by increasing the intensity of fluctuations or by adding the external noise, since it can be considered as an addition to the thermal noise (see Eqs. (3.4) and (3.5)). Then, for $\tau \leq T$, the RS and hysteresis appear (as shown in Fig. 5c and 5d) and with the further increase in noise intensity, the switching time $\tau$ becomes shorter and hysteresis disappears step by step (as shown in Fig. 5e-5j).

The amplitude of $R_m(t)$ variation always grows with noise and reaches some maximum level, when the hysteresis loop size shrinks down to zero. The growing amplitude of memristance $R_m(t)$ variation with noise was reported earlier in [49]. In the present paper, we show that this effect is accompanied by a gradual increase and subsequent decrease in the *I-V* hysteresis loop area.

The corresponding time series of the current response of memristive devices to the driving sinusoidal signal with an amplitude of 1 V and with a minimum intensity of external noise $\theta_\zeta$ = 0.09 mV²s are shown in Fig. 6a and 6b on different time scales. The current jumps corresponding to typical SET and RESET transitions during one sinusoidal period are marked by arrows in Fig. 6b.

On both time scales, it is seen that the switching is irregular. This is due to the fact that not every negative half-cycle of a sine wave leads to a complete SET process, and not every positive half-cycle ends in a complete RESET transition. The current values in the state immediately after RESET are characterized by a larger variation, which is highlighted for clarity by the gray rectangular strip within 10 periods of the sinusoidal signal.

As the noise intensity is increased, the following basic regularities are established. First of all, with an increase in the noise amplitude, the RS becomes regular even for the sub-threshold driving signal, the switching completeness (degree of completion of the SET and RESET processes) increases, which leads to an increase in the ratio of currents (resistances) in extreme resistive states. By analogy with the early theoretical and experimental works [49,50], we can consider this relation as a useful signal in the memristor response and as a measure of the switching quality. Secondly, increasing the noise intensity improves the switching stability, which manifests itself in a significant decrease in the variation of the resistive states after the SET and RESET transitions. Third, exceeding the optimum noise level leads to a deterioration of the switching parameters [53]. The most representative time series showing the revealed regularities corresponding to the noise intensities of 1.37, 14.43, 34.16 mV²s are shown in Fig. 6c – 6h.

Using the experimental *I-V* characteristics partially presented in the left panels of Fig. 5, the average values of current through the memristor were determined in extreme states at the input signal of 0.5 V. Since the current is limited to a compliance value of 300 $\mu$A during the SET transition, the current variation in the HRS after the RESET switching was taken as a measure of switching stability. Both of these quantitative characteristics are presented in Fig. 7, depending on the noise intensity for a constant amplitude of the driving periodic signal of 1 V.

From Fig. 7 it follows that the resistance ratio is non-monotonously dependent on the noise level and has a pronounced maximum for a noise intensity of 14.43 mV²s. An additional maxi- mum at $\theta_\zeta$ =1.37 mV²s is noteworthy and most likely corresponds to the optimum noise level to stabilize the SET transition, which occurs at lower voltages. The variation of resistive states decreases as the noise level increases and reaches saturation when the optimum noise values are reached. Small variation values for noise in- tensities of less than 1.37 mV²s are associated with the incomplete SET process and lower current absolute values in the LRS [53].

Thus, the revealed RS regularities are in qualitative agreement with the classical SR phenomenon, however, careful statistical analysis and modeling are required to support this observation and establish its mechanism.

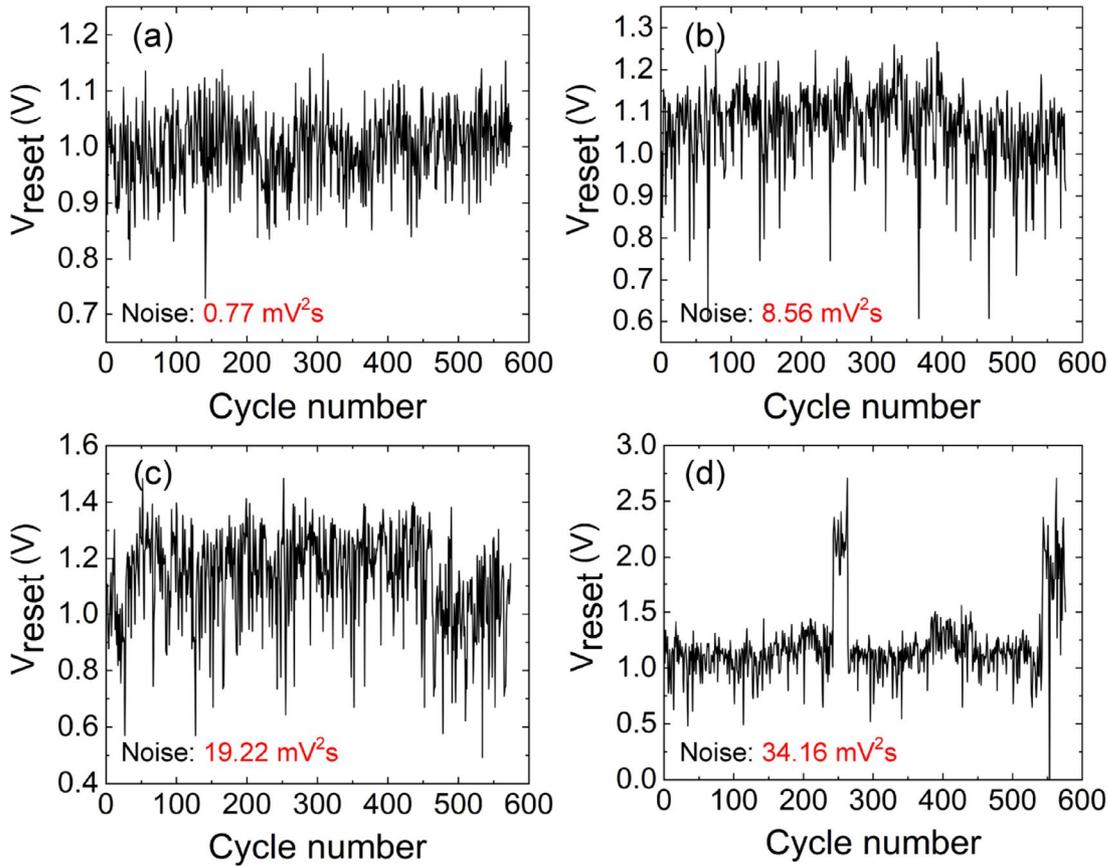

**Fig. 9.** Experimental RESET voltages obtained for each measured cycle for the RS series analyzed at different noise intensities.

*4.2. Time Series Statistical Analysis*

The TSSA has been employed to characterize the statistical features of the memristive device operation variables through a long RS series [54]. From the statistical viewpoint, it can be extracted information related to the correlation of successive RS cycles and the inherent stochasticity of RS.

Considering all the current versus time data in a continuous manner, the time series would produce persistent autocorrelation since the data "inertia" is high. Therefore, we employed other strategies based on data selected on a cycle-to-cycle basis, as reported below. This is the usual procedure to analyze the cycle-to-cycle variability [61] and it is also of great use here.

As a criterion for parameter extraction, we decided to take $I_{reset}$ as the maximum current of each current cycle and $I_{set}$ as the minimum current, both with its corresponding $V_{reset}$ and $V_{set}$ voltages (Fig. 8) [62]. Taking into account the presence of noise, this is a reasonable approach to define a robust parameter extraction methodology. We have only used $V_{reset}$ for our study, in this case, as commented above, the extraction is reasonable.

In the case with the higher noise intensity, we could have voltage values that enter the region that produces RESET or SET for a short time. In this case, we could have different RESET events within a single input voltage sinusoidal half period, and the RS series would be altered as well as the autocorrelation function (ACF) calculation shown below. In most cases, for noise intensities below 34.16 mV$^2$s these voltage spikes do not alter much the CFs since the thermal inertia and the ion migration delays connected to the RS processes avoid a great modification in the CF nature. This assumption is backed by the results of the ACFs shown below. In this respect, from the SR viewpoint, a RRAM device is different from a threshold sensor due to the "inertia" in terms of temperature increase (Joule heating) and ion migration needed for the mechanisms behind RS to go through a SET or RESET process.

In Fig. 9 we show the corresponding extracted RESET voltages for input signals with different noise intensities. The ACFs and partial autocorrelation functions (PACFs) of the data samples are shown in Fig. 10 [54]. The ACF is a function of the number of cycles $k$ and measures the influence/connection between $V_{reset}$ separated by $k$ cycles ($k$ distant lags). In addition, the PACF measures the same correlation but erasing the connection due to the inter-mediated lags (1, 2, …, $k$–1), that is, in the PACF, the dependencies of each two cycles are evaluated making sure that statistical crossed dependencies by means of cycles in between are eliminated.

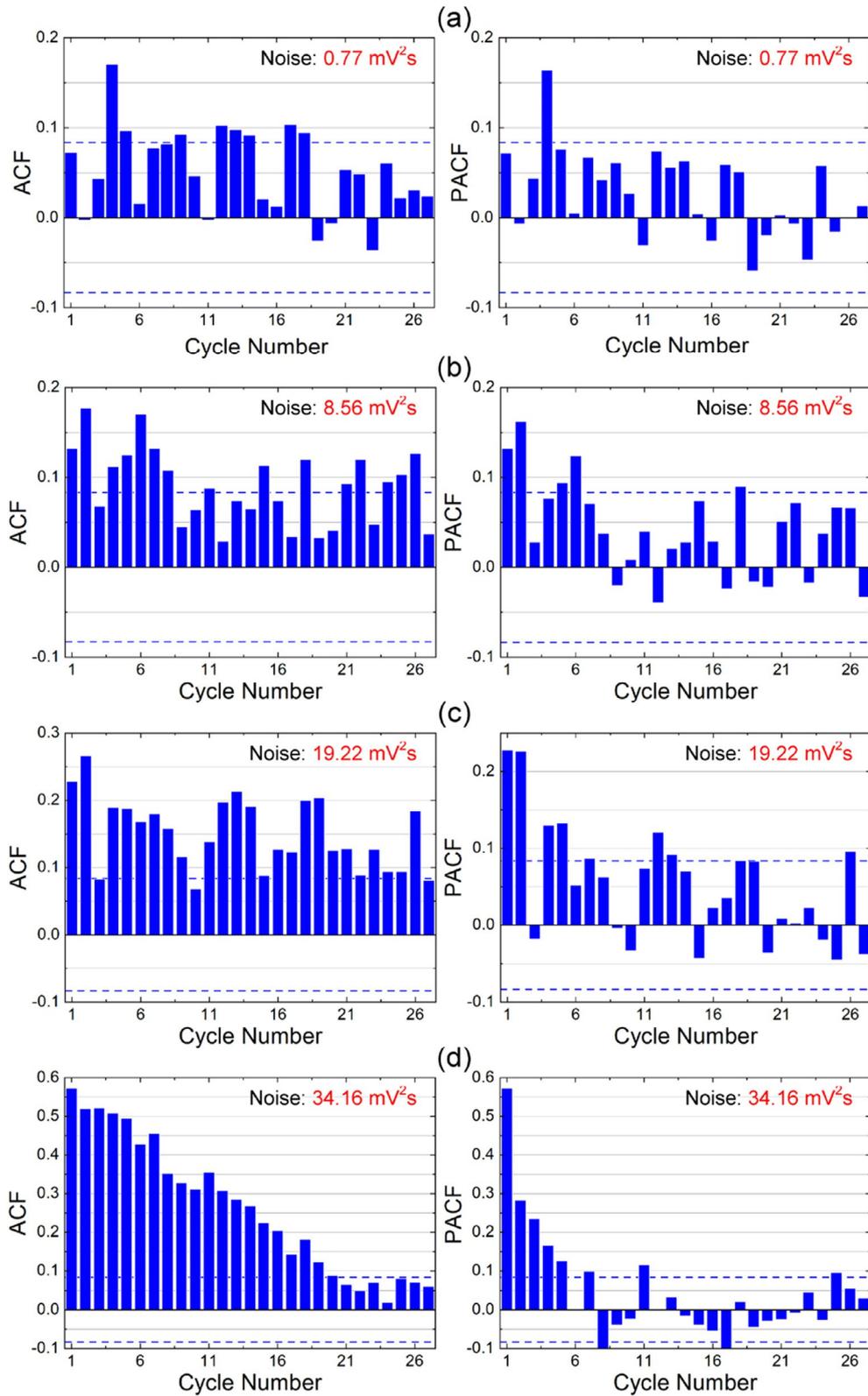

**Fig. 10.** ACF (left panels) and PACF (right panels) vs. cycle lag (distance apart in cycles within a RS series; for a cycle lag 1, the ACF and PACF of consecutive cycles are measured and so on) for the $V_{reset}$ series described in Fig. 9, corresponding to the input signals with different noise intensities. The ACF and PACF minimum and maximum threshold bounds are shown by the dashed lines.

The time series analytical expressions for these data are given below and can be used for the RS series forecasting at different noise intensities. For this purpose, autoregressive (AR), autoregressive and moving average (ARMA) and autoregressive integrated moving average (ARIMA) modeling methodologies are employed as described in the references [54,63,64].

The $V_{reset}$ model proposed for the noise intensity of 0.77 mV²s in the TSSA context is an AR(4) model (some null coefficients were found in the model for $t$-1, $t$-2 and $t$-3)

$$V_{reset,t} = 0.8209 + 0.17\, V_{reset,t-4} + e_t \tag{4.1}$$

where $e_t$ stands for a residual that accounts for the model error [63,64].

The $V_{reset}$ model proposed for the noise intensity of 8.56 mV²s is an ARIMA(0,1,1) and is given as

$$V_{reset,t} = V_{reset,t-1} + e_t - 0.9520\,e_{t-1} \qquad (4.2)$$

In this series, stationarity does not hold [63,64]. In the case of a nonstationary data series, TSSA theory proposes changes of variables that lead the newly derived series to fulfill the stationary requirements that are required prior to the modeling process. ARIMA approaches can be employed instead of the AR or ARMA modeling schemes.

The $V_{reset}$ model proposed for the noise intensity of 19.22 mV$^2$s is based on an ARIMA(1,1,3) approach and is given as

$$V_{reset,t} = 0.52\,V_{reset,t-1} + 0.48\,V_{reset,t-2} + e_t - 0.3777\,e_{t-1} - 0.3409\,e_{t-2} - 0.1845\,e_{t-3} \qquad (4.3)$$

The $V_{reset}$ model proposed for the noise intensity of 34.16 mV$^2$s is an ARIMA(0,1,1) and is given as

$$V_{reset,t} = V_{reset,t-1} + e_t - 0.7204\,e_{t-1} \qquad (4.4)$$

It should be noted that the high noise intensity makes the device to quit the RS cycles marked by the input signal. Therefore, some of the RESET voltages obtained are modified by the noise distortion in the formation/rupture processes of the CFs. In this respect, these features suggest that the operation proposed here should be limited to noise intensities below 34.16 mV2s. This fact described from a different viewpoint (the one linked to TSSA), clarifies other aspects of the SR phenomenon under study here.

The results of the ACFs and PACFs show that for higher noise level the autocorrelation in the $V_{reset}$, calculated as explained below, is higher in general. In this respect, in terms of RS, and in terms of the CFs that are formed in each cycle, the noise + input signal affects the RS inertia. For the highest noise intensities, the spikes linked to this high voltage amplitude noise can change RS cycles and alter the response of the device. These results, from a different perspective, are in line with the conventional relation of the SNR relation versus noise amplitude and the well-known phenomenon that highlights SR.

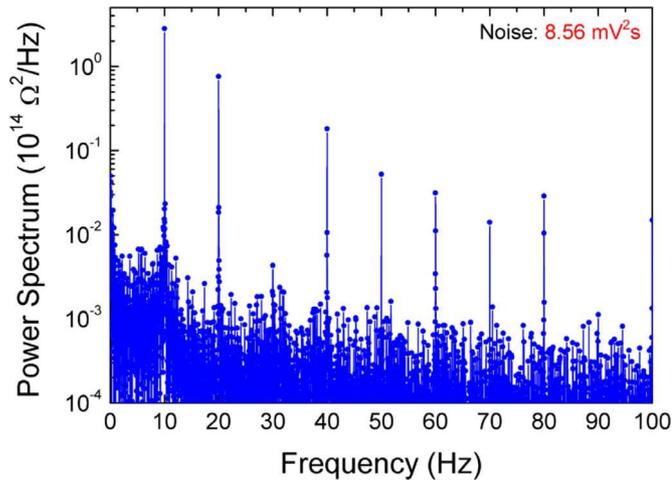

**Fig. 11.** Typical power spectrum of the memristance $R_m(t)$ of device under study measured for a noise intensity of 8.56 mV$^2$s.

*4.3. Spectral Response and SNR*

The most informative parameter of RS is the value of memristor resistance (memristance). As it was described above in the investigated system, the memristance Rm (t) oscillates (switches) appropriately with the periodical driving voltage. According to the coarse-grained stochastic model, the averaged-over-ensemble amplitude of memristance oscillations grows with noise from a null value at zero noise intensity up to a maximum level. This property was pointed out also in [49]. The experimental raw data for a single device are not averaged and therefore they are much noisier, especially for large intensities of external noise. The typical power spectrum of Rm (t) is shown in Fig. 11.

If we consider the SNR measured for the main harmonic at 10 Hz in the spectrum of Rm (t) relative to the level of noise, we obtain the dependence of SNR on external intensity shown in Fig. 12. One can see the dependence of SNR on noise intensity typical for the SR phenomenon. For a very small noise intensity, we observe the most usual dependence – the SNR decreases with external noise. Then, SNR reaches a minimum and starts to grow. The maximum SNR is reached at $\theta_\zeta \approx 10 - 12$ mV2s and then the SNR decreases again. This behavior is very similar to the classical case of SR for symmetric bistable or threshold systems with additive noise [45] as well as to the more sophisticated asymmetric cases with multiplicative and additive noise sources studied later in [65].

In this study, we deliberately limited ourselves to considering only the first harmonic in the spectral response of the memristive device, which is sufficient to demonstrate SR. A more detailed analysis of the power redistribution between the first and higher harmonics is very promising for studying memristive devices under conditions of periodic driving and will be the subject of a separate study. In particular, this is evidenced by recent results [66] showing that hysteresis of memristive systems can be unambiguously distinguished in their Fourier spectrum from the linear or nonlinear response of systems without memory.

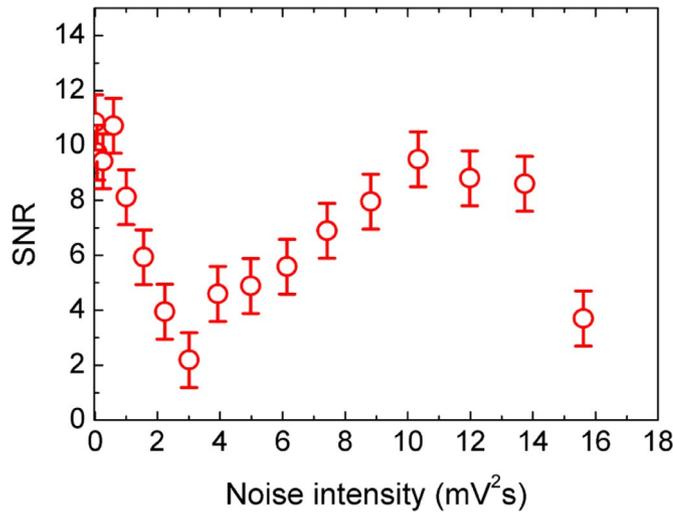
**Fig. 12.** Memristance SNR vs. noise intensity experimentally obtained for the memristive device under study.

## 5. Conclusions

In the present work, we have investigated the main features of RS parameters variation under the application of a Gaussian noise signal added to the driving voltage. It is shown that, according to the coarse-grained stochastic model [34], we may consider the external noise as an addition to the thermal one. Therefore, the variation of external voltage noise intensity is equivalent to the variation of the intensity describing the thermal noise. This is different from the stochastic model based on a direct insertion of an external noise source to the dynamical model proposed in [50], where the resulting noise becomes multiplicative and nonlinear while the thermal noise is absent (not taken into account).

The application of harmonic driving voltage with the fixed amplitude of 1 V and an additive Gaussian noise source with increasing intensity provides the appearance of regular RS, which is observed as a hysteresis loop in I-V characteristics or as oscillations of memristance $R_m(t)$. The size of the hysteresis loop initially grows with noise and then decreases disappearing for a large enough noise, while the amplitude of $R_m(t)$ oscillations are always growing with the noise in average but becoming more and more noisy for each separate realization. The increasing hysteresis loop and amplitude of RS were predicted first in [49] while the further evolution of hysteresis was revealed in this work.

We have also found from a time series statistical analysis that there are two processes going on: a simple one linked to SR where the memristor works as a threshold sensor and another one, connected to the previous one, linked to RS. If RS were an instantaneous process we would have a typical SR case, but it is not the case, that is why we obtain an increase in autocorrelation for $V_{reset}$ with a noise level. Time series analytical expressions are given for the $V_{reset}$ obtained in the measurements and can be used for the RS series forecasting at different noise intensities.

The SNR obtained for the first harmonic of RS value from the $R_m(t)$ power spectrum changes with noise non-monotonically. The SNR increases with the external noise intensity reaching a maxi- mum value and then decreasing. This behavior is a typical manifestation of the SR and constructive role of noise in nonequilibrium memristive systems.

## Acknowledgments

This work was supported by the Government of the Russian Federation (Agr. No. 074-02-2018-330 (2)).

## References

[1] Ielmini D, Waser R. Resistive Switching: From Fundamentals of Nanoionic Re- dox Processes to Memristive Device Applications. Weinheim, Germany: Wi- ley-VCH Verlag GmbH & Co KGaA; 2016.
[2] Lee SH, Zhu X, Lu WD. Nanoscale resistive switching devices for mem- ory and computing applications. Nano Res 2020;13:1228–43. doi:10.1007/ s12274-020-2616-0.
[3] Wang Z, Wu H, Burr GW, Hwang CS, Wang KL, Xia Q, Yang JJ. Resistive switching materials for information processing. Nat Rev Mater 2020;5:173–95. doi:10.1038/s41578-019-0159-3.
[4] Zhang Y, Wang Z, Zhu J, Yang Y, Rao M, Song W, et al. Brain-inspired comput- ing with memristors: Challenges in devices, circuits, and systems. Appl Phys Rev 2020;7:011308. doi:10.1063/1.5124027.
[5] Erokhin V. Memristive Devices for Neuromorphic Applications: Comparative Analysis. Bionanoscience 2020;10:834–47. doi:10.1007/s12668-020-00795-1.
[6] Mikhaylov A, Pimashkin A, Pigareva Y, Gerasimova S, Gryaznov E, Shchanikov S, Zuev A, Talanov M, Lavrov I, Demin V, Erokhin V, Lobov S, Mukhina I, Kazantsev V, Wu H, Spagnolo B. Neurohybrid memristive CMOS- integrated systems for biosensors and neuroprosthetics. Front Neurosci 2020 14:358. doi:10.3389/fnins.2020.00358.
[7] Zhang W, Mazzarello R, Wuttig M, Ma E. Designing crystallization in phase- change materials for universal memory and neuro-inspired computing. Nat Rev Mater 2019;4:150–68. doi:10.1038/s41578-018-0076-x.
[8] Prezioso M, Merrikh-Bayat F, Hoskins BD, Adam GC, Likharev KK, Strukov DB. Training and operation of an integrated neuromorphic network based on metal-oxide memristors. Nature 2015;521:61–4. doi:10.1038/nature14441.
[9] Burr GW, et al. Neuromorphic computing using non-volatile memory. Adv. Phys. X 2016;2:89–124. doi:10.1080/23746149.2016.1259585.
[10] Ielmini D, Wong HSP. In-memory computing with resistive switching devices. Nat Electron 2018;1:333–43. doi:10.1038/s41928-018-0092-2.
[11] Yu S. Neuro-inspired computing with emerging nonvolatile memorys. Proc. IEEE 2018;106:260–85. doi:10.1109/JPROC.2018.2790840.
[12] Zidan MA, Strachan JP, Lu WD. The future of electronics based on memristive systems. Nat Electron 2018;1:22–9. doi:10.1038/s41928-017-0006-8.
[13] Ambrogio S, et al. Equivalent-accuracy accelerated neural-network training us- ing analogue memory. Nature 2018;558:60–7. doi:10.1038/s41586-018-0180-5.
[14] Waser R, Aono M. Nanoionics-based resistive switching memories. Nat Mater 2007;6:833–40. doi:10.1038/nmat2023.
[15] Lanza M, Wong H-SP, Pop E, Ielmini D, Strukov D, Regan BC, Larcher L, Vil- lena MA. Recommended Methods to Study Resistive Switching Devices. Adv Electron Mater 2019;5:1800143. doi:10.1002/aelm.201800143.
[16] Strukov DB, Snider GS, Stewart DR, Williams RS. The missing memristor found. Nature 2008;453:80–3. doi:10.1038/nature06932.
[17] Chua L. Memristor-The missing circuit element. IEEE Trans Circuit Theory 1971;18:507–19. doi:10.1109/TCT.1971.1083337.
[18] Chua LO, Kang Sung Mo. Memristive devices and systems. Proc IEEE 1976;64:209–23. doi:10.1109/PROC.1976.10092.
[19] Ielmini D. Resistive switching memories based on metal oxides: Mecha- nisms, reliability and scaling. Semicond Sci Technol 2016;31:1–25. doi:10.1088/ 0268-

1242/31/6/063002.

[20] Pankratov E L, Spagnolo B. Optimization of impurity profile for *p-n* junction in heterostructures. Eur. Phys. J. B 2005;46:15–19. doi:10.1140/epjb/ e2005-00233-1.
[21] Strukov D B, Alibart F, Williams R S. Thermophoresis/diffusion as a plausible mechanism for unipolar resistive switching in metal–oxide–metal memristors Appl. Phys. A 2012;107:509–18. doi:10.1007/s00339-012-6902-x.
[22] Pérez E, Maldonado D, Acal C, Ruiz-Castro JE, Alonso FJ, Aguilera AM, et al. Analysis of the statistics of device-to-device and cycle-to-cycle variability in TiN/Ti/Al:HfO$_2$/TiN RRAMs. Microelectron Eng 2019;214:104–9. doi:10.1016/j. mee.2019.05.004.
[23] Roldan JB, Maldonado D, Jimenez-Molinos F, Acal C, Ruiz-Castro JE, Aguilera AM, et al. Reversible dielectric breakdown in h-BN stacks: a statistical study of the switching voltages. IEEE Int Reliab Phys Symp 2020:1–5 2020- April. doi:10.1109/IRPS45951.2020.9129147.
[24] Mikhaylov AN, Gryaznov EG, Belov AI, Korolev DS, Sharapov AN, Guseinov DV, Tetelbaum DI, Tikhov SV, Malekhonova NV, Bobrov AI, Pavlov DA, Gerasimova SA, Kazantsev VB, Agudov NV, Dubkov AA, Rosário CMM, Sobolev NA, Spagnolo B. Field- and irradiation-induced phenomena in memristive nano- materials. Phys. Status Solidi C - Current Topics in Solid State Physics 2016;13:870–81. doi:10.1002/pssc.201600083.
[25] Zhu L, Zhou J, Guo Z, Sun Z. An overview of materials issues in resistive random access memory. J Mater 2015;1:285–95. doi:10.1016/j.jmat.2015.07.009.
[26] Trapatseli M, Cortese S, Serb A, Khiat A, Prodromakis T. Impact of ultra- thin Al$_2$O$_{3-y}$ layers on TiO$_{2-x}$ ReRAM switching characteristics. J Appl Phys 2017;121:1–9. doi:10.1063/1.4983006.
[27] Mikhaylov A, Belov A, Korolev D, Antonov I, Kotomina V, Kotina A, et al. Multilayer Metal Oxide Memristive Device with Stabilized Resistive Switching. Adv Mater Technol 2020;5:1900607. doi:10.1002/admt.201900607.
[28] Alibart F, Gao L, Hoskins BD, Strukov DB. High precision tuning of state for memristive devices by adaptable variation-tolerant algorithm. Nanotechnology 2012;23:075201. doi:10.1088/0957-4484/23/7/075201.
[29] Berdan R, Prodromakis T, Toumazou C. High precision analogue memristor state tuning. Electron Lett 2012;48:1105–7. doi:10.1049/el.2012.2295.
[30] Pershin YV, Slipko VA. Dynamical attractors of memristors and their networks. EPL (Europhysics Lett 2019;125:20002. doi:10.1209/0295-5075/125/20002.
[31] Pershin YV, Slipko VA. Bifurcation analysis of a TaO memristor model. J Phys D Appl Phys 2019;52:505304. doi:10.1088/1361-6463/ab4537.
[32] Driscoll T, Pershin Y V, Basov DN, memristor Di Ventra M Chaotic. Appl Phys A 2011;102:885–9. doi:10.1007/s00339-011-6318-z.
[33] Guseinov D, Matyushkin I, Chernyaev N, Mikhaylov A, Pershin Y. Capacitance effects can make memristor chaotic. Chaos, Solitons & Fractals 2021;142:110699. doi:10.1016/j.chaos.2021.110699.
[34] Agudov NV, Safonov AV, Krichigin AV, Kharcheva AA, Dubkov AA, Valenti D, Guseinov DV, Belov AI, Mikhaylov AN, Carollo A, Spagnolo B. Nonstationary distributions and relaxation times in a stochastic model of memristor. J Stat Mech Theory Exp 2020;2020:024003. doi:10.1088/1742-5468/ab684a.
[35] Spagnolo B, Valenti D, Guarcello C, Carollo A, Persano Adorno D, Spezia S, Pizzolato N, Di Paola B. Noise-induced effects in nonlinear relaxation of condensed matter systems. Chaos, Solitons & Fractals 2015;81:412–24. doi:10. 1016/j.chaos.2015.07.023.
[36] Spagnolo B, Guarcello C, Magazzù L, Carollo A, Persano Adorno D, Valenti D. Nonlinear Relaxation Phenomena in Metastable Condensed Matter Systems. Entropy 2017;19:20. doi:10.3390/e19010020.
[37] Valenti D, Magazzù L, Caldara P, Spagnolo B. Stabilization of quantum metastable states by dissipation. Phys Rev B 2015;91:235412. doi:10.1103/ PhysRevB.91.235412.
[38] Dubkov AA, Spagnolo B. Verhulst model with Lévy noise excitation. Eur. Phys. J. B 2008;65:361–7. doi:10.1140/epjb/e2008-00337-0.
[39] Falci G, La Cognata A, Berritta M, D'Arrigo A, Paladino E, Spagnolo B. Design of a Lambda system for population transfer in superconducting nanocircuits. Phys. Rev. B 2013;87:214515. doi:10.1103/PhysRevB.87.214515.
[40] Spagnolo B, Valenti D. Volatility Effects on the Escape Time in Financial Market Models. Int J Bifurc Chaos 2008;18:2775–86. doi:10.1142/S0218127408022007.
[41] Giuffrida A, Valenti D, Ziino G, Spagnolo B, Panebianco A. A stochastic interspecific competition model to predict the behaviour of Listeria monocytogenes in the fermentation process of a traditional Sicilian salami. European Food Research and Technology 2009;228:767–75. doi:10.1007/s00217-008-0988-6.
[42] Denaro G, Valenti D, La Cognata A, Spagnolo B, Bonanno A, Basilone W, Mazzola S, Zgozi S, Aronica S. Spatio-temporal behaviour of the deep chlorophyll maximum in Mediterranean Sea: Development of a stochastic model for pico- phytoplankton dynamics. Ecological Complexity 2013;13:21–34. doi:10.1016/j. ecocom.2012.10.002.
[43] Spagnolo B, Dubkov AA, Pankratov AL, Pankratova EV, Fiasconaro A, Ochab— Marcinek A. Lifetime of metastable states and suppression of noise in Interdisciplinary Physical Models. Acta Phys Pol B 2007;38:1925–50.
[44] Spagnolo B, Dubkov AA, Agudov NV. Enhancement of stability in randomly switching potential with metastable state. The European Physical Journal B 2004;40:273–81. doi:10.1140/epjb/e2004-00268-8.
[45] Gammaitoni L, Hänggi P, Jung P, resonance Marchesoni F Stochastic. Rev Mod Phys 1998;70:223–87. doi:10.1103/RevModPhys.70.223.
[46] Doering CR, Gadoua JC. Resonant activation over a fluctuating barrier. Phys Rev Lett 1992;69:2318–21. doi:10.1103/PhysRevLett.69.2318.
[47] Pizzolato N, Fiasconaro A, Persano Adorno D, Spagnolo B. Resonant activation in polymer translocation: new insights into the escape dynamics of molecules driven by an oscillating field. Physical Biology 2010;7:034001. doi:10.1088/ 1478-3975/7/3/034001.
[48] Guarcello C, Valenti D, Carollo A, Spagnolo B. Effects of Lévy noise on the dynamics of sine-Gordon solitons in long Josephson junctions. Journal of Statistical Mechanics: Theory and Experiment 2016;2016:054012. doi:10.1088/ 1742-5468/2016/05/054012.
[49] Stotland A, Di Ventra M. Stochastic memory: Memory enhancement due to noise. Phys Rev E 2012;85:011116. doi:10.1103/PhysRevE.85.011116.
[50] Patterson GA, Fierens PI, Grosz DF. On the beneficial role of noise in resistive switching. Appl Phys Lett 2013;103:074102. doi:10.1063/1.4819018.
[51] Ntinas V, Rubio A, Sirakoulis GC, Rodriguez R, Nafria M. Experimental Investigation of Memristance Enhancement. IEEE/ACM Int. Symp. Nanoscale Archit. 2019;1:1–2. doi:10.1109/NANOARCH47378.2019.181299.
[52] Ntinas V, Rubio A, Sirakoulis GC, Aguilera ES, Pedro M, Crespo-Yepes A, et al. Power-efficient Noise-Induced Reduction of ReRAM Cell's Temporal Variability Effects. IEEE Trans Circuits Syst II Express Briefs 2020;7747 1–1. doi:10.1109/ TCSII.2020.3026950.
[53] Filatov DO, Vrzheshch DV, Tabakov OV, Novikov AS, Belov AI, Antonov IN, et al. Noise-induced resistive switching in a memristor based on ZrO$_2$ (Y)/Ta$_2$O$_5$ stack. J Stat Mech Theory Exp 2019;2019:124026. doi:10.1088/1742-5468/ ab5704.
[54] Roldán JB, Alonso FJ, Aguilera AM, Maldonado D, Lanza M. Time series statistical analysis: A powerful tool to evaluate the variability of resistive switching memories. J Appl Phys 2019;125:174504. doi:10.1063/1.5079409.
[55] Lanzara E, Mantegna RN, Spagnolo B, Zangara R. Experimental study of a nonlinear system in the presence of noise: The stochastic resonance. Am J Phys 1997;65:341–9. doi:10.1119/1.18520.
[56] McNamara B, Wiesenfeld K, Roy R. Phys Rev Lett 1988;60:2626–9. doi:10.1103/ PhysRevLett.60.2626.
[57] Mantegna RN, Spagnolo B. Stochastic Resonance in a Tunnel Diode in the Presence of White or Colored Noise. Nuovo Cimento D 1995;17:873–81. doi:10. 1007/BF02451845.
[58] Mantegna RN, Spagnolo B, Trapanese M. Linear and Nonlinear Experimental Regimes of Stochastic Resonance. Phys Rev E 2001;63:011101. doi:10.1103/ PhysRevE.63.011101.
[59] Andronov A et al. Proc. II Workshop NATO SfP973799 Semiconductors (Nizhnii Novgorod, Russia, 2002) 2002: 38.
[60] Agudov NV, Devyataykin RV, Malakhov AN. Transient bimodality of nonequilibrium states in monostable systems with noise. Radiophys Quantum Electron 1999;42:902–10. doi:10.1007/BF02677104.
[61] Rodriguez N, Maldonado D, Romero FJ, Alonso FJ, Aguilera AM, Godoy A, et al. Resistive Switching and Charge Transport in Laser-Fabricated Graphene Oxide Memristors: A Time Series and Quantum Point Contact Modeling Approach. Materials (Basel) 2019;12:3734. doi:10.3390/ma12223734.
[62] Villena MA, Roldán JB, Jimenez-Molinos F, Suñé J, Long S, Miranda E, et al. A comprehensive analysis on progressive reset transitions in RRAMs. J Phys D Appl Phys 2014;47:205102. doi:10.1088/0022-3727/47/20/205102.
[63] Brockwell PJ, Davis RA. Introduction to Time Series and Forecasting 2016;92. doi:10.1007/978-3-319-29854-2.
[64] Bisgaard S, Kulahci M. Time Series Analysis and Forecasting by Example. Wiley; 2011. doi:101002/9781118056943.
[65] Qiao Z, Lei Y, Lin J, Niu S. Stochastic resonance subject to multiplicative and additive noise: The influence of potential asymmetries. Phys Rev E 2016;94:052214. doi:10.1103/PhysRevE.94.052214.
[66] Pershin Y V., Chien C-C, Di Ventra M. The Fourier signatures of memristive hysteresis 2020. arXiv:2010.01313.